\RequirePackage{fix-cm}

\documentclass[smallextended]{svjour3}       
\smartqed  
\usepackage{graphicx}
\usepackage{xspace}
\usepackage{color}
\usepackage{dcolumn}
\usepackage{bm}
\usepackage{times}
\usepackage{changes}
\usepackage{hyperref}
\usepackage{amsmath}
\usepackage{amsfonts}

\definecolor{myurlcolor}{rgb}{0,0,0.7}
\definecolor{myrefcolor}{rgb}{0.8,0,0}

\hypersetup{colorlinks, linkcolor=myrefcolor,
citecolor=myurlcolor, urlcolor=myurlcolor}

\def\be{\begin{equation}}
\def\ee{\end{equation}}
\def\ben{\begin{eqnarray}}
\def\een{\end{eqnarray}}
\def\bea{\begin{array}}
\def\eea{\end{array}}
\def\bc{\begin{center}}
\def\ec{\end{center}}
\newcommand{\bei}{\begin{itemize}}
\newcommand{\eei}{\end{itemize}}
\newcommand{\bee}{\begin{enumerate}}
\newcommand{\eee}{\end{enumerate}}

\def\beq{\begin{equation}}
\def\eeq{\end{equation}}

\def\<{\langle}
\def\>{\rangle}

\begin{document}

%
%
%

\journalname{Foundations of Physics}

\title{Communication strength of correlations
violating monogamy relations}



\author{Waldemar K\l{}obus      \and
        Micha\l{} Oszmaniec \and
				Remigiusz Augusiak \and
				Andrzej Grudka 
}


\institute{Waldemar K\l{}obus  \and Andrzej Grudka
\at Faculty of Physics, Adam Mickiewicz University, 61-614 Pozna\'n, Poland \\
\and Micha\l{} Oszmaniec \and Remigiusz Augusiak
\at ICFO--Institut de Ciencies Fotoniques, Mediterranean Technology Park \\
08860 Castelldefels (Barcelona), Spain
\and Micha\l{} Oszmaniec
\at Center for Theoretical Physics, Polish Academy of Sciences, Warszawa, Poland \\
\email{oszmaniec@cft.edu.pl}
}

\date{Received: date / Accepted: date}

\maketitle

\begin{abstract}
In any theory satisfying the no-signaling principle correlations generated
among spatially separated parties in a Bell-type experiment are subject to
certain constraints known as monogamy relations. Recently, in the context of the black hole information loss problem it was suggested that these monogamy relations might be violated.
This in turn implies that correlations arising in such a scenario must violate the no-signaling principle and hence can be used to send classical information between parties. Here, we study the amount of
information that can be sent using such correlations.
To this aim, we first provide a framework associating them with
classical channels whose capacities are then used to quantify
the usefulness of these correlations in sending information.
Finally, we determine the minimal amount of information that can be sent using
signaling correlations violating the monogamy relation associated to the
chained Bell inequalities.


\end{abstract}

\section{Introduction}

In recent years a lot of research has been devoted to probabilistic
nonsignaling theories \cite{GNT,GNT2}. They are formulated in terms of
\textit{boxes}, that is, families of probability distributions
describing correlations generated in a Bell-type experiment by spatially
separated observers. The boxes are required to satisfy the no-signaling
principle which means that expectation values seen by some of the observers
cannot depend on the measurement choices made by the remaining ones  (see e.g. Ref.
\cite{review}). A particular example of a theory obeying the no-signaling
principle is quantum mechanics. It was realized, however, that there exist
nonsignaling theories which lead to higher violations of Bell inequalities than
it is allowed by quantum mechanics \cite{PR}.
This discovery raised a debate as to whether such supra-quantum nonsignaling
correlations can be found in Nature (see, e.g., Refs. \cite{principles}).

One of the most interesting features of the nonsignaling correlations is that
they are monogamous \cite{Toner,Marcin,monog,Ravi}. Consider for instance
a three-partite scenario in which Alice and Bob violate the
Clauser-Horne-Shimony-Holt (CHSH) \cite{CHSH} or the chained \cite{BC90}
Bell inequality  up to its maximal algebraic value. Then, each of
Alice's or Bob's observables appearing in it cannot be correlated with an
arbitrary observable measured by Eve \cite{monog}. This fact found important
applications in cryptography based on nonsignaling principle \cite{KD} and
randomness amplification \cite{RA}---tasks that are impossible in classical
world.

Although all well-established physical theories satisfy the no-signaling principle, there is at least one important physical phenomenon where monogamy relations can be violated---the black hole information loss problem. It was argued by Almheiri, Marolf, Polchinski, and Sully that if information escapes from black hole, then one can check if the entanglement monogamy is violated \cite[page 5]{Almheiri}. Later, Oppenheim and Unruh showed that by performing measurements on three particles in a ``polygamous entangled state'' near black hole, one can send superluminal signals, thus giving rise to a box violating the no-signaling principle.
%
%
Let us also note that if one allows for post-selection in the Bell-type experiment, violation of monogamy relations can appear in quantum mechanics \cite{Preskill}, which can have applications to black hole information loss problem \cite{Horowitz}.

Let us now consider a box violating a monogamy relation.
Then, this box must be signaling. Then, the natural question appears:
\textit{how can the box be used to send information from some parties to the
other parties, and, moreover, how much communication can be sent?} In this
Letter we answer these questions for three-partite boxes 
which violate monogamy relations for the CHSH and the chained Bell inequalities. We
also present a very simple proof of the monogamy relations introduced in
Ref. \cite{monog}. By putting monogamy relations in a broader framework allows one
to get a better understanding of their structure.

Before presenting our results, we need to
introduce some notation and terminology. Imagine that three parties $A$, $B$,
and $E$ perform a Bell-type experiment in which $A$ and $B$
can measure one of $M$ observables, denoted $A_i$ and $B_j$, respectively,
while the external observer $E$ measures a single observable, which
we also denote by $E$. We assume that all these observables have two outcomes
$\pm 1$, denoted $a$, $b$, and $e$. The correlations that are generated in such
an experiment are described by a set of probabilities
$\{p(A_iB_jE)\equiv p(a,b,e|A_i,B_j)\}$, where $p(a,b,e|A_i,B_j)$ is the
probability of obtaining $a,b,e$ when $A_i$, $B_j$ and $E$ have been measured by
$A$, $B$, and $E$, respectively. In what follows we arrange these
probabilities in vectors denoted $\vec{p}$ and refer to them as
\textit{boxes}. We then say that the distribution $\{p(A_iB_jE)\}$
obeys the no-signaling principle (it is \textit{nonsignaling}) if all of its
marginals describing a subset of parties is independent of the measurement
choices made by the remaining parties, i.e., 
\ben
\sum_{a}p(a,b,e|A_i B_j) &=& \sum_{a}p(a,b,e|A_k B_j), 
\een
and
\ben
\sum_{b}p(a,b,e|A_i B_j) &=& \sum_{b}p(a,b,e|A_i B_k),
\een
are satisfied for any triple $i,j,k$.
Then, by $\langle XY\rangle_Z$ we denote
the standard bipartite expectation value of the product of observables $X$ and $Y$, which in general might be conditioned on the third party's measurement choice $Z$. An example of such conditional expectation value is
\begin{equation}
\langle A_iE\rangle_{B_j}=\sum_{a,b,e=\pm1} a\cdot	e\cdot p(a,b,e|A_i,B_j).
\end{equation}
If $\{p(a,b,e|A_i,B_j)\}$ is
nonsignaling, then clearly $\langle XY\rangle_{Z}=\langle XY\rangle_{Z'}\equiv\langle XY\rangle$ for
any choice of $X$, $Y$, and $Z\neq Z'$.

Let us finish the introductory section be defining what we mean by "classical information" in the Bell-type scenario introduced above. In this scenario, the parties have access only to measurements of classical random variables associated to observables\footnote{For simplicity we consider only the situation in which the observables have two outcomes, but this is not a serious restriction.}  $A_i$, $B_j$ and $E$. In this sense the results of the experiments are inherently classical and carry the classical information to which we refer in latter parts of the paper. For instance, if $A$ decides to measure the observable $A_i$, its result is either $-1$ or $1$ and thus can be encoded in one logical bit. Analogously, a result of the joint measurement of the observables $A_i$,$B_j$ and $E$ is can be encoded in tree logical bits. 

\section{A simple derivation of a monogamy relation for the CHSH Bell
inequality}

For clarity we begin our considerations with the simplest scenario of $M=2$.

The key ingredient of our framework is a simple proof of the
monogamy relation obeyed by any nonsignaling
probability distribution $\{p(a,b,e|A_i,B_j)\}$ \cite{Toner,monog}:
\begin{equation}\label{monrel}
|I_{AB}|+2|\langle B_0 E\rangle|\leq 4,
\end{equation}
where $I_{AB}$ stands for the
Bell expression giving rise to the well-known
CHSH Bell inequality \cite{CHSH}
\begin{equation}\label{CHSH}
I_{AB}:=\langle A_0 B_0 \rangle + \langle A_1 B_0 \rangle + \langle A_1 B_1 \rangle - \langle A_0 B_1 \rangle\leq 2.
\end{equation}
The inequality (\ref{monrel}) compares the nonlocality shared by $A$ and $B$, as
measured by the violation of (\ref{CHSH}), to the (classical) correlations that
the external party $E$ can establish with the outcomes of $B_0$. It should be
noticed that it remains valid if in the last correlator, $B_0$ is replaced by
any $A_i$ or $B_i$ (for clarity, however, we proceed with a fixed measurement
$B_0$). Also, without any loss of generality we can assume that both $I_{AB}$
and $\langle B_0E\rangle$ are positive; if this is not the case, we redefine
observables $A_0$, $A_1$ and/or $E$ in the following way: $A_0 \rightarrow
-A_0$, $A_1 \rightarrow -A_1$, and/or $E \rightarrow -E$. Consequently, in what
follows we omit the absolute values in (\ref{monrel}).

In order to prove (\ref{monrel}), let us first make the following observation.
Suppose that for some random variables $X$, $Y$ and $Z$ taking values $\pm1$
there exists the joint probability distribution $p(XYZ)$. Then, the latter
fulfils the following inequalities
\begin{eqnarray}\label{basicIneq}
(-1)^i\langle X Y \rangle_Z + (-1)^j\langle Y Z \rangle_X +(-1)^k \langle
X Z \rangle_Y \leq 1, \label{lem3}
\end{eqnarray}
with $i,j,k=0,1$ such that $i\oplus j\oplus k=1$, where addition is modulo
2. To prove (\ref{basicIneq}), it suffices to check it for the extremal values of correlators.

Now, one notices that each triple of observables $A_i$, $B_j$ and $E$
is jointly measurable and therefore, for any pair $i,j$, there exists
the joint probability distribution $p(A_iB_jE)$ which must satisfy (\ref{lem3}).
This gives rise to the following four inequalities
\begin{eqnarray}
 \langle A_0 B_0 \rangle_{E} + \langle B_0 E \rangle_{A_0} - \langle A_0 E \rangle_{B_0}& \leq &1, \label{nierownosci1}\\
 \langle A_1 B_0 \rangle_{E} + \langle B_0 E \rangle_{A_1} - \langle A_1 E \rangle_{B_0} &\leq &1, \label{nierownosci2}\\
 \langle A_1 B_1 \rangle_{E} - \langle B_1 E \rangle_{A_1} + \langle A_1 E \rangle_{B_1}& \leq &1,
 \label{nierownosci3}\\
 -\langle A_0 B_1 \rangle_{E} + \langle B_1 E \rangle_{A_0} + \langle A_0 E \rangle_{B_1} &\leq & 1. \label{nierownosci4}
\end{eqnarray}
By summing these up and using the fact that in a nonsignaling theory $\langle X Y \rangle_{Z}=\langle X Y \rangle_{Z'}$ for any $Z\neq Z'$, one obtains \eqref{monrel}.

\section{Signaling boxes as classical channels}

Let us assume that correlators $\langle B_0 E \rangle_{A_0}$ and $\langle B_0 E \rangle_{A_1}$ are equal (later we will
show how this assumption can be relaxed). Then the monogamy relation
\eqref{monrel} is well defined. It bounds the possible correlations achievable
in any no-signaling theory between outcomes of measurements performed by the
three parties $A$, $B$ and $E$. If it is violated by some probability
distribution $\vec{p}$, then the latter must violate the no-signalling principle, in which case we call such a box \textit{signaling}. In
other words, if $\vec{p}$ violates (\ref{monrel}), then values of some
bipartite correlators become dependent on the measurement choice made by the
third party. This dependence allows one to use such signaling boxes to send
information from a single party to the remaining two parties. To illustrate this
idea, suppose that a box $\vec{p}$ violates the relation \eqref{monrel} by
$\Delta>0$, that is,
\begin{equation}
R(\vec{p})\equiv I_{AB}+2\langle B_0E\rangle=4+\Delta.
\end{equation}
Then, by adding the inequalities \eqref{nierownosci1}-\eqref{nierownosci4}, one
concludes that
\begin{equation}
\langle A_0 E \rangle_{B_0} -  \langle A_0 E \rangle_{B_1} + \langle A_1 E \rangle_{B_0} - \langle A_1 E \rangle_{B_1}
 + \langle B_1 E \rangle_{A_1} - \langle B_1 E \rangle_{A_0} \geq \Delta.
\label{ineqcor}
\end{equation}
Consequently, in at least one of the three pairs
\begin{eqnarray}
S_{B\to AE}^0&= &\{ \langle A_0 E \rangle_{B_{0}}, \langle A_0 E \rangle_{B_{1}}\}, \label{set1}\\
S_{B\to AE}^1&= &\{ \langle A_1 E \rangle_{B_{0}}, \langle A_1 E \rangle_{B_{1}}\}, \label{set2}\\
S_{A\to BE}^1&= &\{\langle B_1 E \rangle_{A_1},  \langle B_1 E \rangle_{A_0}\}, \label{set3}
\end{eqnarray}
the correlators must differ. In particular, in one of them the
difference must not be lower than $\Delta/3$. The correlators in $S_{B\to
AE}^i$ correspond to signaling from $B$ to the pair $A$ and $E$, while those in
$S_{A\to BE}^i$ to signaling from $A$ to $B$ and $E$.

\begin{figure}[h]
\begin{center}
  \includegraphics[width=0.5\textwidth]{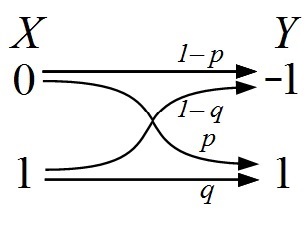}
  \caption{A binary classical channel that can be associated to one of the pair
of correlators
  $S^i_{A\to BE}$ and $S^i_{B\to AE}$ with $i=0,1$.
  The random variable $X\in\{0,1\}$ corresponds to the choice of measurement by
$A$ or $B$, while the random variable $Y\in\{-1,1\}$ represents the product of
either $A_i$ and $E$ or $B_i$ and $E$, depending on the considered pair of
correlators. Finally, $p=p(Y=1|X=0)$ and $q=p(Y=1|X=1)$ are the transition
probabilities defining the channel.}
  \label{ch}
	\end{center}
\end{figure}

Let us now assume, without any loss of generality, that
\ben\label{piq}
\langle A_0 E \rangle_{B_0} -  \langle A_0 E \rangle_{B_1} >0,
\een
which can be rewritten as
$p-q >0$, where $p\equiv p(A_0E=1|B_0)$ and $q\equiv p(A_0E=1|B_1)$.
It then follows from (\ref{piq}) that the probability that the parties
$A$ and $E$ obtain the same results while measuring $A_0$ and $E$,
respectively, depends on whether the remaining party measures $B_0$ or
$B_1$. This gives rise to a binary asymmetric channel, denoted
$\mathcal{C}_{B\to AE}^0$, with the input and output alphabets
$\{B_0,B_1\}$ and $\{-1,1\}$, respectively, and the transition probabilities
given by (see Fig. \ref{ch})
\begin{eqnarray}\label{Ensanche}
\hspace{-0.5cm}&p(A_0E=1|B_0) = p, \quad
p(A_0E=-1|B_0) = 1-p,& \\
\hspace{-0.5cm}&p(A_0E=1|B_1) = q, \quad
p(A_0E=-1|B_1) = 1-q. &\label{Ensanche2}
\end{eqnarray}

Importantly, the above reasoning opens the possibility to quantify the 
communication strength of boxes violating the no-signalling principle 
by  the concept of classical channel capacity. This is a standard notion in classical information theory which,  according to Shannon's  noisy-channel coding theorem 
, quantifies the amount of information that a classical channel can transmit per single use \cite{shannon2}.
In particular, the capacity of a 
binary asymmetric channel with the transition probabilities \eqref{Ensanche}-\eqref{Ensanche2}
can be explicitly written as \cite{moser}:
\begin{equation}\label{cepq}
C(p,q) = \frac{pH(q)-qH(p)}{q-p}+ \log_2 \left( 1 + 2^{\frac{H(p)-H(q)}{q-p}} \right)
\end{equation}
with $H(p)$ being the standard binary entropy.

Analogously, one associates classical channels to the other two pairs of
correlators $S_{B\to AE}^1$ and $S_{A\to BE}^1$. As a result, any box violating
the monogamy relation (\ref{monrel}) gives rise to three
channels $\mathcal{C}_{A\to BE}^1$ and
$\mathcal{C}_{B\to AE}^i$ of capacities

\begin{equation}
C_{A\to BE}^1=C(p_A^1,q_A^1)\ \ \text{and}\ \  C_{B\to AE}^i=C(p_B^i,q_B^i)\ ,
\end{equation}
where
\begin{equation}
p_X^i=(1+x_X^i)/2\ \ \text{and}\ \  q_X^i=(1+y_X^i)/2
\end{equation}
are probabilities
corresponding to the correlators
\begin{equation}
x_B^{i}=\left\langle A_{i}E\right\rangle
_{B_0}\ \ \text{and}\ \  y_B^{i}=\langle A_iE\rangle_{B_1}
\end{equation}
for $i=0,1$, and $x_A^1=\langle B_1E\rangle_{A_1}$ and $y_A^1=\langle B_1E\rangle_{A_0}$.

It should finally be noticed that a box violating (\ref{monrel})
might also feature signaling from one or two parties
to a single one; still, by definition, $E$ cannot signal to $A$ and $B$.
Such situations could, however, make our considerations difficult
to handle and in order to avoid them, in what follows we restrict our attention
to a subclass of boxes whose all one-partite expectation values $\langle X\rangle_{YZ}$
with $X,Y,Z=A_i,B_j,E$ are zero. Let us stress, nevertheless, that
this assumption does not influence at all what we have said so far,
as, for any box violating (\ref{monrel}), there exists another one with exactly the same
two-body correlators (and giving rise to exactly the same channels and the same violation of (\ref{monrel}))
whose all one- and three-partite expectation values vanish. Precisely,
given a probability distribution $\{p(A_iB_jE)\}$, the box $\{p'(A_iB_jE)\}$
with
\begin{equation}
p'(A_iB_jE)=\frac{1}{2}[p(A_iB_jE)+p(\bar{A_i}\bar{B_j}\bar{E})],
\end{equation}
where $\bar{A}_i=-A_i$ etc., has the same two-body correlators as
$\{p(A_iB_jE)\}$
and all its one-partite and three-partite mean values are zero. Below
we then restrict our attention to boxes having only bipartite correlators
non-vanishing. They form a convex set denoted by $\mathcal{P}$. Let
also $\mathcal{P}_{\Delta}$
be the subset of $\mathcal{P}$ composed of boxes $\vec{p}$ for which
$R(\vec{p})=4+\Delta$ with $\Delta\in[0,2]$.

\section{Communication strength of boxes violating monogamy
relation (\ref{monrel})}
Our aim now is to explore the communication strength of boxes violating
(\ref{monrel}) in terms of capacities of the three associated channels.
To this aim, we will first determine a set of constraints on elements
of $\mathcal{P}_{\Delta}$ that fully characterizes correlators giving rise to
these channels. It follows from \eqref{lem3} that
$p(A_0B_1E)$ and $p(A_1B_1E)$ obey the following inequalities
\begin{eqnarray}
\langle A_1 B_1 \rangle_{E} + \langle B_1 E \rangle_{A_1} - \langle A_1 E \rangle_{B_1} &\leq & 1, \label{nierownosci3a}\\
-\langle A_0 E \rangle_{B_1} - \langle B_1 E \rangle_{A_0} - \langle A_0 B_1 \rangle_{E} &\leq & 1. \label{nierownosci4a}
\end{eqnarray}
Replacing Eqns. \eqref{nierownosci3} and \eqref{nierownosci4} with Eqns. \eqref{nierownosci3a} and \eqref{nierownosci4a}, we obtain four non-equivalent sets of four inequalities of the form (\ref{nierownosci1})-(\ref{nierownosci4}).
By adding them in each of these sets and assuming that (\ref{monrel}) is
violated by $\Delta\in\left(0,2\right]$, we arrive at the following inequalities
\ben
 x_B^0 - y_B^0 + x_B^1 - y_B^1 + x_A^1 - y_A^1 &\geq &\Delta, \label{niert1}\\
 x_B^0 + y_B^0 + x_B^1 - y_B^1 - x_A^1 - y_A^1 &\geq &\Delta, \label{niert2}\\
 x_B^0 - y_B^0 + x_B^1 + y_B^1 + x_A^1 + y_A^1 &\geq &\Delta, \label{niert3}\\
 x_B^0 + y_B^0 + x_B^1 + y_B^1 - x_A^1 + y_A^1 &\geq &\Delta. \label{niert4}
\een
In Appendix \hyperref[AppA]{A} we show that for a given $\Delta\in[0,2]$, these
inequalities and the trivial conditions $-1\leq \langle XY
\rangle_Z \leq 1$ are the only restrictions on the two-partite correlators
$x_A^1$, $y_A^1$, $x_B^0$, $y_B^0$, $x_B^1$, and $y_B^1$, which for further
purposes we arrange in a vector $\vec{c}$. In other words, for
any vector of correlators $\vec{c}$ satisfying \eqref{niert1}-\eqref{niert4}
there always exists a probability distribution $\vec{p}$ that realizes
$\vec{c}$ and violates (\ref{monrel}) by $\Delta$. On the level of
correlators, this observation gives us a complete characterization of signaling
in boxes violating the monogamy (\ref{monrel}).

Having the above constraints, we are now in position to study the communication
properties of boxes violating (\ref{monrel}). More precisely, we will determine
the minimal (nonzero) amount of information that can be sent from at least one
party to the remaining two parties using a box $\vec{p}$ such that
$R(\vec{p})=4+\Delta$. We notice that for a given $0<\Delta<2$ one might find a
box for which, e.g., $C_{B\to AE}^0> 0$, $C_{B\to AE}^1=0$ and $C_{A\to BE}^1=0$, and, at the same
time, there exists a box for which $C_{A\to BE}^1> 0$, $C_{B\to AE}^0=0$ and $C_{B\to AE}^1=0$,
yet they both give rise to the same violation of (\ref{monrel}).
For this reason we consider the following quantity that depends on the three
capacities
\ben\label{Cdefprim}
C_\Delta = \min_{\mathcal{P}_{\Delta}} \max \{C_{A\to BE}^1,C_{B\to
AE}^0,C_{B\to AE}^1\},
\een
where, due to what has been previously said, the minimization over
$\mathcal{P}_{\Delta}$ can be replaced by a minimization over
the polytope $\mathcal{Q}_{\Delta}$ of all vectors $\vec{c}$ satisfying
\eqref{niert1}-\eqref{niert4} and the trivial conditions $-1\leq\langle
XY\rangle_Z\leq 1$. The quantity $C_{\Delta}$ tells us that at
least one of the three associated channels to any box from
$\mathcal{P}_{\Delta}$ has capacity at least $C_{\Delta}$.

Clearly, $C_0=0$ and in the case when (\ref{monrel})
is violated maximally, i.e., for $\Delta=2$, $C_{2}$ can be computed almost
by hand and amounts to $C_2=0.158$ (see Appendix \hyperref[AppB]{B}).
For all the intermediate values $0<\Delta<2$ the problem of determining
$C_{\Delta}$ becomes difficult to handle analytically. Still it
can be efficiently computed numerically. This is because
the capacity (\ref{cepq}) is a convex function in both arguments \cite{shannon} and so is
the function $\max \{C_{A\to BE}^1,C_{B\to AE}^0,C_{B\to AE}^1\}$ due to the
well-known property that a function resulting from a pointwise maximization of convex functions is also convex.
Then, the minimization in (\ref{Cdefprim}) is executed over a convex polytope.

\begin{figure}[h]
\begin{center}
 \includegraphics[width=0.5\textwidth]{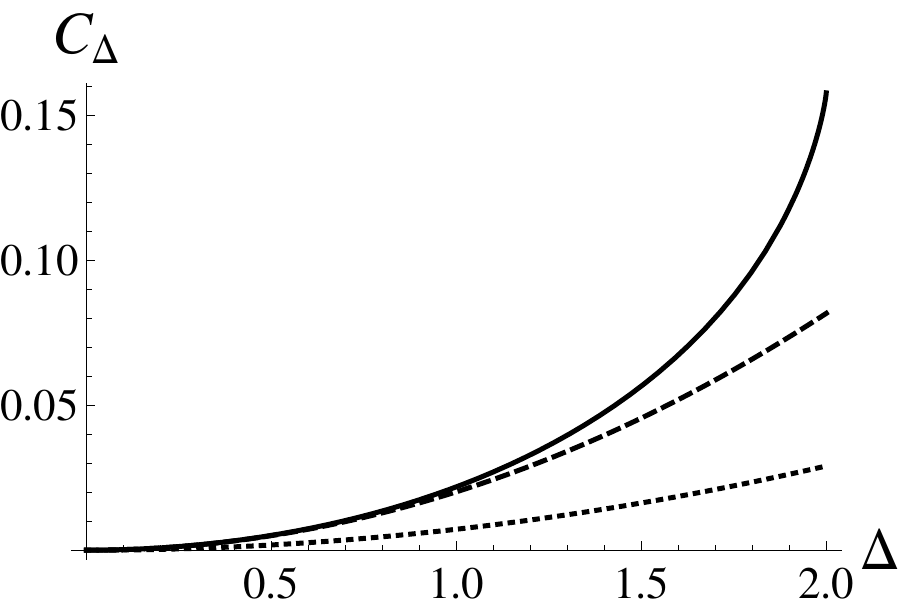}
  \caption{The communication strength $C_{\Delta}$ of signaling
  boxes violating the monogamy relation (\ref{monrel}) by $\Delta$ as a function
of $\Delta$ (solid line). As expected, $C_{\Delta}$ grows with $\Delta$, that
is, the higher the violation of (\ref{monrel}) the more information can be sent
through the associated channels. For comparison, we also present the lower bound
on $C_{\Delta}^M$ given by (\ref{Gava}) for $M=2$ (dashed line) and $M=3$
(dotted line).}
  \label{Cd}
	\end{center}
\end{figure}

The results of our numerical computations are plotted in Fig. \ref{Cd}.
We find that the obtained values of $C_\Delta$ can be realized by boxes
obeying the conditions
$x_B^0 = x_B^1=\Delta/2$ and $x_A^1= - y_A^1=y_B^0=y_B^1$,
and the value of the remaining free parameter $x_A^1$
is set by the condition
\begin{equation}
C((1+\Delta/2)/2,(1+x_A^1)/2) = C((1+x_A^1)/2,(1-x_A^1)/2).
\end{equation}
An exemplary box $\{p(A_iB_jE)\}$ realizing $C_{\Delta}$ and satisfying the
above conditions is given by
\begin{eqnarray}\label{box}
p(A_iB_jE) &=&\frac{1}{4} \left[  1 + A_i E \left( \frac{\Delta}{2}
\delta_{j,0} +
x_A^1 \delta_{j,1} \right) \right]\nonumber\\
&&\times ( \delta_{ij,0} \delta_{A_i B_j ,0} + \delta_{ij,1} \delta_{A_i B_j ,-1} ),
\end{eqnarray}
where $\delta_{m,n}$ denotes the Kronecker delta, and $x_A^1$ is the solution
of the above equation. One can see that for this box all one-partite and
three-partite expectation values vanish. Moreover, its restriction
$\{p(A_iB_j)\}$ to the parties $A$ and $B$ is equivalent to the so-called
Popescu-Rohrlich box \cite{PR}.

Let us conclude by noting that one can also drop the assumption
that the correlators $\<B_0E\>_{A_0}$ and $\<B_0E\>_{A_1}$ are
equal, in which case the monogamy relation \eqref{monrel} reads
\begin{equation}\label{monrel2}
|I_{AB}|+|\langle B_0 E\rangle_{A_0}+\langle B_0 E\rangle_{A_1}|\leq 4.
\end{equation}
Then, following the above methodology one can associate another classical
channel to the pair $S_{A\to
BE}^0=\{\<B_0E\>_{A_0},\<B_0E\>_{A_1}\}$. Our numerics shows, however, that an
addition of this channel in the definition of $C_{\Delta}$ does not change its
value; in particular, the box (\ref{box}) realizes $C_{\Delta}$
and has the property that $\<B_0E\>_{A_0}=\<B_0E\>_{A_1}$.

\section{Generalizing to the chained Bell inequality}
The above considerations
can be applied to a monogamy relation for the generalization of
the CHSH Bell inequality to any number of dichotomic measurements at both
sites---the chained Bell inequality \cite{BC90}.
To recall the latter and the corresponding monogamy, let us assume that now
$A$ and $B$ have $M$ dichotomic measurements at their disposal
denoted $A_k$ and $B_k$ $(k=0,\ldots,M-1)$. The chained Bell inequality
reads \cite{BC90}:
\begin{equation}\label{chained}
I^{M}_{AB}:=\sum_{k=0}^{M-1}(\langle A_kB_k\rangle+\langle A_{k+1}B_{k}\rangle)\leq 2M-2,
\end{equation}
where we use the convention that $A_{M}=-A_0$. As shown in Ref. \cite{monog}, it obeys the following
simple monogamy relation for any nonsignaling correlations
\begin{equation}\label{chainedEve}
|I_{AB}^{M}|+ 2 |\<B_0 E  \>|\leq 2M,
\end{equation}
where $E$ stands for Eve's measurement. As before, we can assume that both $I_{AB}^M$ and $\langle
B_0E\rangle$ are nonnegative, and hence, in what follows
we omit the absolute values in (\ref{chainedEve}).

To proceed with our considerations we first note that analogously to
(\ref{monrel}),
the monogamy (\ref{chainedEve}) can be derived from
(\ref{basicIneq}). Precisely, as any three observables
$A_i$, $B_j$ and $E$ are jointly measurable, due to (\ref{lem3}) the following
set of
$2M$ inequalities
\begin{eqnarray}
&&\label{nier1}\langle A_0B_0\rangle_{E}+\langle B_0E\rangle_{A_0}-\langle A_0E\rangle_{B_0}\leq 1,\\
&&\label{nier2}\langle A_1B_0\rangle_{E}+\langle B_0E\rangle_{A_1}-\langle A_1E\rangle_{B_0}\leq 1
\end{eqnarray}
and
\begin{eqnarray}\label{nier3}
\langle A_{i+j}B_i\rangle_E-(-1)^j\langle B_iE\rangle_{A_{i+j}}+(-1)^j\langle A_{i+j}E\rangle_{B_i}\leq 1
\end{eqnarray}
with $i=1,\ldots,M-1$ and $j=0,1$ must hold. By adding them
and assuming that the no-signaling principle is fulfilled,
one obtains (\ref{chainedEve}).

It is of importance to point out that the inequalities
(\ref{nier1}), (\ref{nier2}), and (\ref{nier3})
form a unique minimal set of inequalities of the form (\ref{lem3}) that, via
the above proof, give rise to the monogamy (\ref{chainedEve}). To be more
explicit,
note that any such set must consists of at least $2M$ inequalities
because there are that many different correlators in the Bell inequality
(\ref{chained}). Then, each of these correlators must appear in any such
$2M$-element set with the same sign as in (\ref{chained}). As one directly
checks, this is enough to conclude that the only $2M$-element set is the one
given in Ineqs. (\ref{nier1})-(\ref{nier3}).

Let us now assume as before that all correlators appearing in the monogamy
relation (\ref{chainedEve}) do not depend on the the third party's
measurements, in particular, $\langle B_0E\rangle_{A_i}=\langle
B_0E\rangle_{A_j}$ for any $i\neq j$. Let then
$\mathcal{P}_{\Delta}^M$ be the convex set of
boxes for which $R_M(\vec{p})=2M+\Delta$ with $\Delta\in[0,2]$.
If $\Delta>0$ there must be some signaling between $A$, $B$, and $E$ in a box
$\vec{p}\in \mathcal{P}_{\Delta}^M$. In particular,
it follows from (\ref{nier1}), (\ref{nier2}), and
(\ref{nier3}) that
\begin{equation}\label{Olot}
\sum_{i=1}^{M-1}(x_A^i-y_A^i)+\sum_{i=1}^{M-1}(x_B^i-y_B^i)
+x_B^0-y_B^0\geq \Delta,
\end{equation}
where
\begin{equation}
x_A^i=\langle B_iE\rangle_{A_i},\ y_A^i=\langle B_iE\rangle_{A_{i+1}},\ x_B^i=\langle A_iE\rangle_{B_{i-1}},\ y_B^i=\langle A_iE\rangle_{B_i},
\end{equation}
and finally
\begin{equation}
x_B^0=\langle A_0E \rangle_{B_0},\ \ \text{and}\ \ y_B^0=\langle A_0E\rangle_{B_{M-1}}.
\end{equation}
Since $\Delta>0$, this implies that in some of the following $2M-1$ pairs
(perhaps all)
\begin{equation}
S^{i}_{A\to BE}=\{x_A^i,y_A^i\},
\end{equation}
with $i=1,\ldots,M-1$, and
\begin{equation}
S^{i}_{B\to AE}=\{x_B^i,y_B^i\},
\end{equation}
with $i=0,\ldots,M-1$ and $B_{-1}\equiv B_{M-1}$,
the correlators must differ. Recall that for the nonsignaling correlations,
correlators belonging to each
$S^{i}_{A\to BE}$ or $S^{i}_{B\to AE}$ are equal.
In the first case this means that
there is signaling from $A$ to $BE$, while in the second one,
from $B$ to $AE$.

Now, analogously to the case $M=2$,
to each pair of correlators $S_{A\to BE}^i$ and $S_{B\to AE}^i$,
can be associated a binary classical channel
of capacity $C(p_A^i,q_A^i)$ and $C(p_B^i,q_B^i)$, respectively,
where $p_{X}^i=(1+x_{X}^i)/2$ and $q_X^i=(1+y_X^i)/2$ with $X=A,B$.
We then quantify the communication strength of boxes from
$\mathcal{P}_{\Delta}^M$ by
\begin{equation}\label{definition2}
C_{\Delta}^M=\min_{\mathcal{P}_{\Delta}^M}\max_{i=1,\ldots,M-1}
\{C(p_A^i,q_A^i),C(p_B^0,q_B^0),C(p_B^i,q_B^i)\},
\end{equation}
which for $M=2$ reduces to $C_{\Delta}$.

Similarly to the case $M=2$, (\ref{Olot}) is not the only inequality
bounding the values of the above correlators. In fact,
each of $2(M-1)$ inequalities in (\ref{nier3}) remains satisfied if the
signs in front of the second and the third correlator are
swapped. By concatenating such swaps, one obtains $4^{M-1}$ sets of $2M$
inequalities and each set when summed up produces an analogous inequality to
(\ref{Olot}). All the resulting inequalities read
\begin{eqnarray}\label{ElPrat}
&&\sum_{i=1}^{M-1}(-1)^{a_i}(x_A^i-y_B^i)+
\sum_{i=1}^{M-2}(-1)^{b_i}(x_B^{i+1}-y_A^i)\nonumber\\
&&\hspace{1cm}+(-1)^c(y_A^{M-1}
+y_B^0)+x_B^1+x_B^0\geq \Delta,
\end{eqnarray}
with $a_i,b_i,c\in\{0,1\}$ for $i=1,\ldots,M-1$. Although we cannot prove
it as in the case $M=2$, we conjecture that all possible values of the
correlators in $S_{A\to BE}^i$ and $S_{B\to AE}^i$ that satisfy inequalities
(\ref{ElPrat}) can always be realized with some signaling probability
distribution $\vec{p}$ for which $R_M(\vec{p})=2M+\Delta$. In general,
by minimizing
\begin{equation}
 \max_{i=1,\ldots,M-1}
\{C(p_A^i,q_A^i),C(p_B^0,q_B^0),C(p_B^i,q_B^i)\}
\end{equation}
over the correlators
$x_A^i$, $y_A^i$, $x_A^i$ and $y_A^i$ satisfying (\ref{ElPrat}) and the trivial
conditions $-1\leq \langle XY\rangle_Z \leq 1$ instead of
$\mathcal{P}_{\Delta}^M$ leads to a lower bound on $C_{\Delta}^M$.

In general, it is a hard task to
compute $C^M_{\Delta}$. Still, one can easily bound it from below by
noting that $C_{\Delta}^M$ majorizes any of the capacities appearing in
(\ref{definition2}). Moreover, by consulting (\ref{Olot}), one finds that at
least one pair in $S_{A\to BE}^i$ or $S_{B\to AE}^i$, say $S^0_{B\to AE}$,
satisfies
\begin{equation}
x_B^0-y_B^0\geq \Delta/(2M-1).
\end{equation}
In terms of probabilities this reads
\begin{equation}
p^0_B - q^0_B \geq \Delta/(4M-2).
\end{equation}
Now, the lower bound on $C_{\Delta}^M$ is
given by the minimum of $C(p^0_B,q^0_B)$ given the above constraint on $p_B^0$
and $q_B^0$.
Using \eqref{cepq} we
conclude that for a given value of $\Delta$, the capacity
attains the minimum for $p^0_B = 1 - q^0_B$, for which the corresponding
binary channel becomes symmetric whose capacity reads $1-H(p_B^0)$. Therefore,
$C(p^0_B,q^0_B)$ is minimized by
\begin{equation}
p^0_B =[1+ \Delta/(4M-2)]/2 \ \ \text{and}\ \ q^0_B =[1- \Delta/(4M-2)]/2,
\end{equation}
which leads to
\ben \label{Gava}
C_{\Delta}^M\geq 1-H\big((1+\tfrac{\Delta}{4M-2})/2\big).
\een
For large $M$ the above lower bound tends to zero and for $M=2$ and $M=3$ it
is plotted in Fig. \ref{Cd}.

\section{Conclusion}

In this work we have shown how signaling correlations
violating monogamy relations could be utilized to send classical information
between space-like separated observers. We have also proposed a quantity
that allows one to quantify the communication strength of such boxes. Moreover, we presented an alternative proof of certain monogamy relations  based on the CHSH \eqref{monrel} and the chained Bell inequalities \eqref{chainedEve}, which contrary to the previous ones allows to understand how the no-signaling principle constraints correlations obtained in a Bell-type experiment.  On the other hand, our results give some insight into the structure of signaling in correlations that are
not monogamous. In particular we showed that from the violation of these monogamy relations one  can infer only about signaling of one party (say Alice) to the remaining two parties participating in the Bell scenario (Bob and Eve). 

Let us finally notice that our analysis suggests that there is
some trade-off between capacities of the three
introduced channels $\mathcal{C}_{B\to AE}^i$ and
$\mathcal{C}_{A\to BE}^1$. Namely, one can satisfy Ineqs.
(\ref{niert1})-(\ref{niert4}) with a signaling box for which
two channels are of zero capacities, but then the third
capacity must be high. In order to lower it, it is
necessary to increase the capacity of one of the two remaining channels.
It seems interesting to determine an analytical relation linking these
capacities.

\begin{acknowledgements}

We thank M. Horodecki, R. Horodecki, P. Kurzy\'{n}ski, M. Lewenstein, J. {\L}odyga and
A. W\'{o}jcik for helpful discussions.  W. K. and A. G. were supported by the
Polish Ministry of Science and Higher Education Grant no. IdP2011 000361. M. O.
was supported by the ERC Advanced Grant QOLAPS, START scholarship granted by Foundation for Polish
Science and the Polish National Science
Centre grant under Contract No. DEC-2011/01/M/ST2/00379. R. A. was supported by
the ERC Advanced Grant OSYRIS, the EU IP SIQS, the John Templeton
Foundation, the Spanish Ministry project FOQUS (FIS2013-46768) and the Generalitat de Catalunya project 2014 SGR
874. W. K. thanks the Foundation of Adam Mickiewicz University in Pozna\'{n} for the
support from its scholarship programme. A.G.
thanks ICFO--Institut de Ci\`{e}ncies Fot\`{o}niques for hospitality.
\end{acknowledgements}




\section*{Appendix A: Conditions for correlators from the violation of monogamy relation (\ref{monrel})}
\label{AppA}

We will now prove that for a particular violation $\Delta$, the inequalities
(\ref{niert1})-(\ref{niert4}) along with the trivial conditions
\begin{equation}\label{trivial}
 -1\leq\langle XY\rangle\leq 1
\end{equation}
satisfied by any pair $X,Y$ constitute the only restrictions on the two-body
correlators $\vec{c}=(x_A^1,y_A^1,x_B^0,y_B^0,x_B^1,y_B^1)$ in the sense that
for any such $\vec{c}$ satisfying inequalities (\ref{niert1})-(\ref{niert4}),
there is a box
$\{p(A_iB_jE)\}$ realizing these correlators and violating the
monogamy $(\ref{monrel})$ by
$\Delta$.


Before passing to the proof, let us first introduce some additional notions. Let
again $B$ be the convex set of all tripartite boxes $\{p(A_iB_jE)\}$ whose all
one and three-partite expectation values vanish. Notice that such boxes are
fully characterized by twelve two-body correlators $\langle A_iB_j\rangle_E$,
$\langle A_iE\rangle_{B_j}$, and $\langle B_jE\rangle_{A_i}$ with $i,j=0,1$,
that is,
\begin{eqnarray}
p(abe|A_iB_jE)&=& \frac{1}{8}(1+ab\langle A_iE\rangle_{B_j}+ae\langle
A_iB_j\rangle_E\nonumber\\
&&\hspace{0.5cm}+be\langle B_jE\rangle_{A_i}),
\end{eqnarray}
for every $a,b,e$ and $i,j$. For further benefits we also arrange the above
expectation values in a vector $\vec{p}$.

Let then $\mathcal{P}$ be a subset of $B$ consisting of boxes for which
the value of the right-hand side of (\ref{monrel}) is
$M(\vec{p})\in[4,6]$, i.e., elements of $\mathcal{P}$ either saturate
the monogamy relation (\ref{monrel}) or violate it. Moreover, by
$\mathcal{P}_{\Delta}$ we denote those
elements of $\mathcal{P}$ for which the value $M(\vec{p})$ is precisely
$4+\Delta$, i.e.,
\begin{equation}
 \mathcal{P}_{\Delta}=\left\{ \vec{p}\in\mathcal{P}|
M(\vec{p})=4+\Delta\right\}.
\end{equation}
Clearly, $\mathcal{P}$ and $\mathcal{P}_{\Delta}$ are polytopes whose vertices
can easily be found, and, in particular, the vertices of $\mathcal{P}$ belong to
either $\mathcal{P}_0$ or $\mathcal{P}_2$.

Let finally $\phi:B\to \mathbb{R}^6$ be a vector-valued function associating
a vector of six correlators $\vec{c}=(x_A^1,y_A^1,x_B^0,y_B^0,x_B^1,y_B^1)$ to
any element of $B$. With the aid of this mapping we can associate to
$P_{\Delta}$ the following polytope
\begin{equation}
 \mathcal{Q}_\Delta=\{(\phi(\vec{p}),\Delta)\in\mathbb{R}^7\,|\,\vec{p}
\in\mathcal{P}_{\Delta}
 \}.
\end{equation}
On the other hand, let us introduce the polytope
$\widetilde{\mathcal{Q}}_{\Delta}$
%
%
of vectors of the form $(\vec{c},\Delta)$
with $\vec{c}$ satisfying the inequalities
(\ref{niert1})-(\ref{niert4}) for some fixed $\Delta$
along with the trivial conditions (\ref{trivial}). By definition,
$\mathcal{Q}_{\Delta}\subseteq\widetilde{\mathcal{Q}}_{\Delta}$ for any
$\Delta$ and our aim now is to prove that
$\mathcal{Q}_{\Delta}=\widetilde{\mathcal{Q}}_{\Delta}$. In
particular, we want to show that any $\vec{c}\in\mathcal{Q}_{\Delta}$ with some
fixed $\Delta\geq 0$ can always be completed to a full probability
distribution $\vec{p}\in\mathcal{P}_{\Delta}$ violating (\ref{monrel}) by
$\Delta$.

With the above goal we define two additional polytopes
\begin{equation}
 \mathcal{Q}_{v}=\{(\phi(\vec{p}),M(\vec{p})-4)\in\mathbb{R}^{7}\,|\,\vec{p}
\in\mathcal{P}\},
\end{equation}
and
\begin{equation}
\widetilde{\mathcal{Q}}_{v}=\bigcup_{\Delta\in[0,2]}\widetilde{\mathcal{Q}}_{
\Delta}.
\end{equation}
Direct numerical computation shows that, analogously to $\mathcal{P}$, the
vertices of $\mathcal{Q}_v$
belong to either $\mathcal{Q}_0$ or $\mathcal{Q}_2$. In the same way one shows
that the vertices of both polytopes $\mathcal{Q}_{v}$ and
$\widetilde{\mathcal{Q}}_{v}$
overlap, which implies that $\mathcal{Q}_v=\widetilde{\mathcal{Q}}_{v}$.
Using then the definition of these sets and the fact that the mapping
$\vec{p}\to (\phi(\vec{p}),M(\vec{p}))$ is linear, one obtains that
$\mathcal{Q}_{\Delta}=\widetilde{\mathcal{Q}}_{\Delta}$ for any $\Delta$.

\section*{Appendix B: Analytical computation of $C_2$}
\label{AppB}

Here we determine analytically the capacity $C_{\Delta}$ in the case when the
monogamy relation (\ref{monrel}) is violated maximally, i.e., for $\Delta=2$.
From Ineqs. \eqref{niert1}-\eqref{niert4} it immediately follows that
$x_B^0=x_B^1=1$, $y_B^0=x_A^1$, and
$y_B^1=-y_A^1$, and the problem of determining $C_2$ considerably simplifies to
\begin{equation}
C_2=\min_{-1\leq \alpha,\beta\leq
1}\max\{\widetilde{C}(1,\alpha),\widetilde{C}(1,\beta),\widetilde{C}(\alpha,
-\beta)\},
\end{equation}
where we have substituted $y_B^0=\alpha$ and $y_B^1=\beta$ and have denoted
$\widetilde{C}(\alpha,\beta)=C((1+x)/2,(1+y)/2)$ with $C$ defined in Eq.
(\ref{cepq}). To compute the above, it is useful to
notice that the function $\widetilde{C}$ satisfies
$\widetilde{C}(\alpha,\beta)=\widetilde{C}(\alpha,\beta)
=\widetilde{C}(-\alpha,-\beta)$, and that it is convex in both arguments
(cf. Ref. \cite{shannon}). The latter implies in particular
that for any $\alpha\leq0$, $\widetilde{C}(1,\alpha)\geq
\widetilde{C}(\alpha,\beta)$
and also $\widetilde{C}(1,\alpha)\geq \widetilde{C}(\alpha,-\beta)$ with $-1\leq
\beta\leq 1$. This observation suggests dividing the square $-1\leq
\alpha,\beta\leq 1$ into four ones (closed) whose facets are given by $\alpha=0$
and $\beta=0$, and determining $C_2$ in each of them.
In fact, whenever $\alpha\leq 0$ or $\beta\leq 0$,
\begin{equation}
C_2=\min_{\alpha,\beta}\max\{\widetilde{C}(1,\alpha),\widetilde{C}(1,\beta)\},
\end{equation}
and by direct checking
one obtains $C_2=0.322$. In order to find $C_2$ in the last region given by
$\alpha\geq 0$ and $\beta\geq 0$,
one first notices $\widetilde{C}(1,\alpha)\geq \widetilde{C}(1,\beta)$ if, and
only if $\alpha\leq \beta$. This, along with the fact that
$\widetilde{C}(\alpha,-\beta)=\widetilde{C}(-\beta,\alpha)=
\widetilde{C}(\beta,-\alpha)$ means that
we can restrict our attention to the case $\alpha\leq \beta$, for which
\begin{equation}
C_2=\min_{\alpha\leq
\beta}\max\{\widetilde{C}(1,\alpha),\widetilde{C}(\alpha,-\beta)\}.
\end{equation}
In the last step
we notice that for any $0\leq \beta\leq 1$, $\widetilde{C}(\alpha,-\beta)$ and
$\widetilde{C}(1,\alpha)$ are, respectively, monotonically increasing and
decreasing functions of $\alpha$.
Additionally, for any $0 \leq \alpha\leq 1$, $\widetilde{C}(\alpha,-\beta)$ is a
monotonically increasing
function of $\beta$. Then, for $\alpha=1$, $\widetilde{C}(1,-1)=1$, while
$\widetilde{C}(1,1)=0$
(recall that we assume that $\alpha\leq \beta$),
and for $\alpha=0$, $\min_{\beta\geq 0}\widetilde{C}(\alpha,-\beta)=0$ and
$\widetilde{C}(1,0)>0$. All this means that
both functions $\widetilde{C}(1,\alpha)$ and $\widetilde{C}(\alpha,-\beta)$
intersect, implying that
$C_2$ lies on the line given by
$\widetilde{C}(1,\alpha)=\widetilde{C}(\alpha,-\beta)$. Finally,
as already mentioned, $\widetilde{C}(\alpha,-\beta)$ is a monotonically
decreasing function of $\beta$ which together with $\alpha\leq \beta$ means that
$\alpha=\beta$ has to be taken. One then arrives at the condition that
$\widetilde{C}(1,\alpha)=\widetilde{C}(\alpha,-\alpha)$, which has a solution
when for $\alpha=0.469$ giving $C_2=0.158$.
By comparing both minima, we finally obtain that $C_2=0.158$.

\end{document}